\documentclass[aps,prl,twocolumn,superscriptaddress,nolongbibliography,showpacs]{revtex4-2}

\usepackage{graphicx}
\usepackage{dcolumn}
\usepackage{bm}
\usepackage{mhchem}
\usepackage{xcolor}

\begin{document}


\title{Resonant Inelastic X-ray Scattering from Electronic Excitations in $\alpha$-RuCl$_3$ Nanolayers}

\author{Zichen Yang}
\affiliation{Max Planck Institute for Solid State Research, Heisenbergstrasse 1, D-70569 Stuttgart, Germany}
\author{Lichen Wang}
\affiliation{Max Planck Institute for Solid State Research, Heisenbergstrasse 1, D-70569 Stuttgart, Germany}
\author{Dong Zhao}
\affiliation{Max Planck Institute for Solid State Research, Heisenbergstrasse 1, D-70569 Stuttgart, Germany}
\author{Mingdi Luo}
\affiliation{Max Planck Institute for Solid State Research, Heisenbergstrasse 1, D-70569 Stuttgart, Germany}
\author{Sourav Laha}
\altaffiliation{Present address: Department of Chemistry, National Institute of Technology Durgapur, Mahatma Gandhi Avenue, Durgapur-713209, India}
\affiliation{Max Planck Institute for Solid State Research, Heisenbergstrasse 1, D-70569 Stuttgart, Germany}
\author{Achim G\"uth}
\affiliation{Max Planck Institute for Solid State Research, Heisenbergstrasse 1, D-70569 Stuttgart, Germany}
\author{Takashi Taniguchi}
\affiliation{National Institute for Materials Science, 1-1 Namiki, Tsukuba, 305-0044, Japan}
\author{Kenji Watanabe}
\affiliation{National Institute for Materials Science, 1-1 Namiki, Tsukuba, 305-0044, Japan}
\author{Bettina V. Lotsch}
\affiliation{Max Planck Institute for Solid State Research, Heisenbergstrasse 1, D-70569 Stuttgart, Germany}
\author{Jurgen H. Smet}
\affiliation{Max Planck Institute for Solid State Research, Heisenbergstrasse 1, D-70569 Stuttgart, Germany}
\author{Matteo Minola}
\affiliation{Max Planck Institute for Solid State Research, Heisenbergstrasse 1, D-70569 Stuttgart, Germany}
\author{Hlynur Gretarsson}
\affiliation{Deutsches Elektronen-Synchrotron DESY, Notkestraße 85, D-22607 Hamburg, Germany}
\author{Bernhard Keimer}
\email{B.Keimer@fkf.mpg.de}
\affiliation{Max Planck Institute for Solid State Research, Heisenbergstrasse 1, D-70569 Stuttgart, Germany}

\date{\today}

\begin{abstract}
We present Ru $L_3$-edge resonant inelastic x-ray scattering (RIXS) measurements of spin-orbit and {\it d-d} excitations in exfoliated nanolayers of the Kitaev spin-liquid candidate RuCl$_3$. Whereas the spin-orbit excitations are independent of thickness, we observe a pronounced red-shift and broadening of the d-d excitations in layers with thickness below $\sim$7 nm. Aided by model calculations, we attribute these effects to distortions of the RuCl$_6$ octahedra near the surface. Our study paves the way towards RIXS investigations of electronic excitations in various other 2D materials and heterostructures.
\end{abstract}

\maketitle

Since the discovery of the Scotch-tape exfoliation method, \cite{graphene_science_Geim,graphene_PNAS_Geim} two-dimensional (2D) materials and heterostructures have grown into a unique laboratory for quantum physics. By reconfiguring the crystal symmetry and reducing the dimensionality of the electron system, exfoliation of atomically thin sheets can generate electronic ground states with physical properties radically different from those of bulk analogues. Superstructures generated by vertical stacking  \cite{2D_heterostructure_device_review,2D_heterostructure_science_review,TMDC_review} and lateral twisting \cite{graphene_magicangle_nature} of these sheets add numerous options for control and design of collective quantum phenomena. To realize these perspectives, experimental information on the electron-electron and electron-lattice interactions that determine the stability of different quantum states is indispensable. Research on bulk quantum materials has shown that data from energy- and momentum-resolved spectroscopic probes provide particularly insightful information for realistic model calculations. Prominent examples include angle-resolved photoemission spectroscopy (ARPES) and inelastic neutron scattering (INS), which yield the dispersion relations of electronic bands and collective excitations, respectively. Whereas ARPES has been widely applied to 2D materials, however, INS experiments are not feasible because they require sample volumes in the cm$^{3}$ range.

Resonant Inelastic X-ray Scattering (RIXS) has recently gained prominence as a momentum-resolved spectroscopic probe of electronic and vibrational excitations. \cite{RIXS_review_RMP_1,RIXS_review_RMP_2} Whereas the energy resolution of RIXS for collective magnetic and vibrational excitations remains lower than the one of INS, the latest generation of RIXS instruments has enabled detection of such excitations in many materials, and RIXS additionally probes charge and orbital excitations over a wide spectral range (meV to eV). This includes ligand field excitations which are hard to access with other spectroscopic techniques and whose knowledge is often crucial to understand the physics of 2D materials and VdW heterostructures. Crucially, the large resonant enhancement of the scattering cross section at X-ray absorption edges, combined with the high photon flux at modern synchrotron sources, endow RIXS with a sensitivity that greatly exceeds the one of INS and has allowed the detection of excitations from microcrystals and monolayer-thick films. In exfoliated layers and van-der-Waals (VdW) heterostructures, RIXS has the potential to reveal a wealth of information about atomic-scale interactions including crystalline electric fields, spin-orbit coupling, magnetic exchange, and electron-phonon interactions. The element-selective nature of RIXS allows one to focus exclusively on the properties of a specific layer of a VdW heterostructure, without interference from substrates and protective capping layers. However, as the lateral dimensions of typical exfoliated nanoflakes are below the x-ray beam diameter, such experiments present formidable challenges, and the potential of RIXS for research on 2D materials remains largely untapped.

\begin{figure}[!htb]
	\centering
	\includegraphics{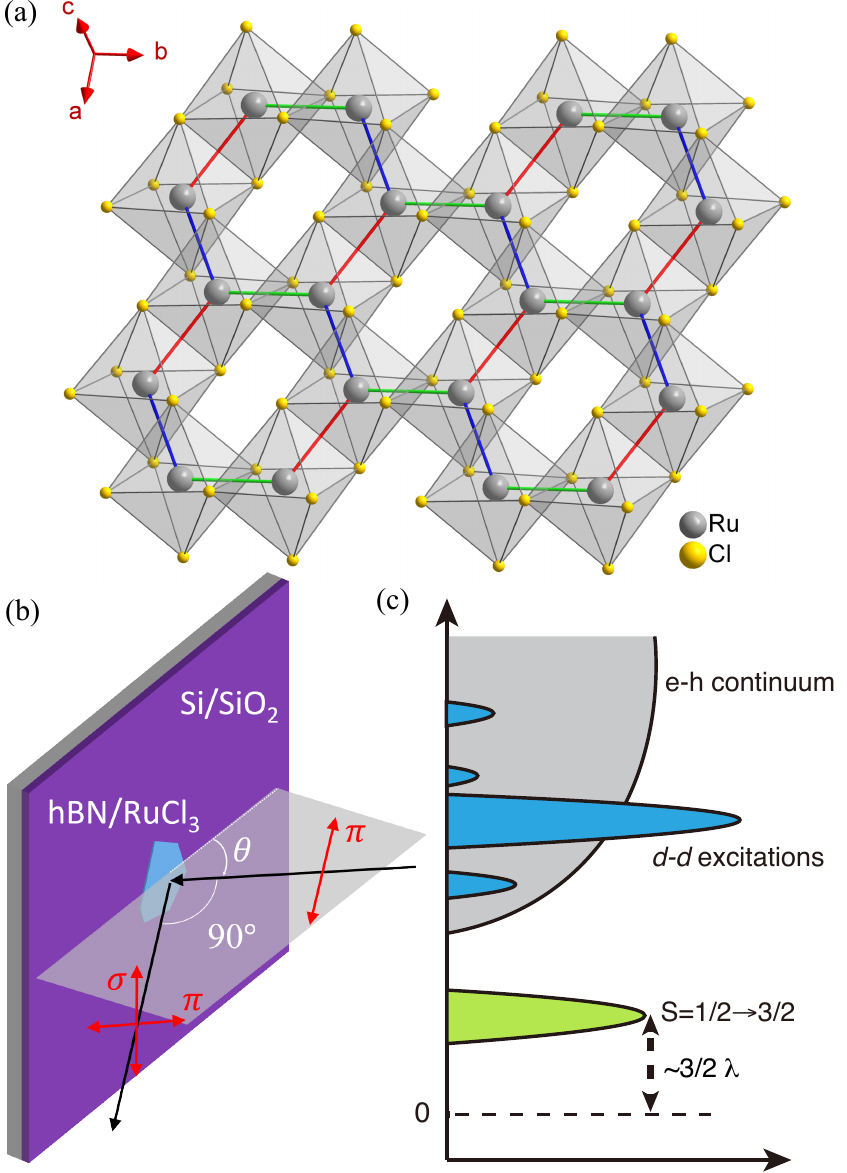}
	\caption{(a) In-plane crystal structure of \ce{RuCl3}. Red, green and blue lines illustrate the bond-directional Kitaev interactions between magnetic Ru ions on the honeycomb lattice. (b) Schematic of the scattering geometry. The incident x-ray photons are $\pi$-polarized, and the polarization of the scattered x-ray photons is not analyzed. The scattering angle is fixed at 90$^{\circ}$ throughout the experiment to suppress charge scattering. (c) Schematic of the elementary excitations of \ce{RuCl3}. The $\tilde{S}=1/2 \rightarrow 3/2$ spin-orbit exciton (green) is located at the excitation energy $\sim 3/2\lambda$. The higher-energy {\it d-d} excitations (blue) are superposed by the electron-hole continuum (grey). }
	\label{fig:Fig1}
\end{figure}

Here we report RIXS experiments on exfoliated nanolayers of $\alpha$-\ce{RuCl3} (\ce{RuCl3} hereafter), a possible solid-state realization of the intensely investigated Kitaev spin liquid  \cite{Kitaev,spin_liquid_Balents_nature,Kitaev_SOC_theory}. The crystal structure of \ce{RuCl3} [Fig.\ref{fig:Fig1}(a)] is composed of edge-sharing RuCl$_6$ octahedra with magnetic Ru atoms arranged on a honeycomb lattice. As a consequence of the strong spin-orbit coupling (SOC) of Ru, the low-energy magnetic dynamics can be described in terms of pseudospins $\tilde{S}=1/2$ that interact via bond-directional, frustrated Kitaev interactions as well as conventional Heisenberg and off-diagonal exchange interactions. The confluence of these interactions drives the system into a state with zigzag antiferromagnetic order at low temperatures. Nevertheless, a continuum of (possibly fractionalized) magnetic excitations \cite{RuCl3_raman_Sandilands,RuCl3_INS_naturemat} and a magnetic-field-induced phase with highly unusual thermal transport properties \cite{RuCl3_torque_PRL,RuCl3_Matsuda_nature,RuCl3_Matsuda_science,RuCl3_Ong_naturephysics} have been ascribed to Kitaev interactions. Since adjacent honeycomb layers are chemically bonded predominatly through van-der-Waals forces, \ce{RuCl3} has also been investigated in the form of exfoliated nanosheets \cite{RuCl3_Raman_JPCS,RuCl3_Raman_IOP,RuCl3_Raman_PRB,RuCl3_transport_Raman_KK} and VdW heterostructures \cite{RuCl3_Gr_strain_PRL,RuCl3_Gr_PRB,RuCl3_Gr_transport_Jurgen,RuCl3_Gr_Raman_ACS}. These developments raise the prospect of studying magnetism in the 2D limit, without the influence of the interlayer interactions that are found to have a non-negligible influence on the magnetic structure of bulk RuCl$_3$\cite{RuCl3_interlayer_dispersion_PRB_2019,RuCl3_interlayer_theory_PRB_2020}. They also open up perspectives for targeted modification of the electronic properties, for instance by doping charge carriers into the correlated pseudospin system via doping across heterointerfaces, or by interfacial proximity coupling to other quantum states such as superconductivity \cite{SC_Kitaev_1,SC_Kitaev_2,SC_Kitaev_3,SC_Kitaev_4}.

Motivated by these prospects and by the detailed information on crystal-field, spin-orbit, and exchange interactions obtained from previous RIXS experiments on bulk RuCl$_3$, \cite{Suzuki2021} we prepared a series of RuCl$_3$ nanoflakes of varying thickness down to 3.5 nm and lateral dimensions comparable to those of the x-ray beams required for RIXS. We obtained high-quality Ru $L_3$-edge RIXS spectra on all samples, without any sign of X-ray beam damage. With decreasing thickness, we observed a systematic red-shift and broadening of electronic transitions from the Ru $t_{2g}$ orbitals in the crystal-field ground-state into excited states in the $e_g$ manifold, whereas intra-atomic spin-orbit excitations are thickness independent. Based on ionic model calculations and comparison to prior surface-sensitive studies, we attribute this trend to an altered ligand field near the surface, which controls the ratio of Kitaev and Heisenberg interactions and hence the magnetic ground state. Our results indicate that RIXS experiments on a variety of 2D materials and VdW heterostructures -- and the resulting wellspring of information on electronic interactions -- are within reach of current instrumentation. 

The experiments were performed at the intermediate X-ray energy RIXS spectrometer (IRIXS) at the Dynamics Beamline P01 of the synchrotron PETRA III, DESY  \cite{SrRuO_Hakuto_naturemat,Suzuki2021,CaRuO_Hlynur_RIXS_PRB,Hlynur_RIXS_JSR,CaRuO_Joel_RIXS_PRB}, which operates at the Ru-$L_3$ absorption edge (photon energy 2837 eV). We used IRIXS in two configurations, i.e. with inline high-resolution monochromator (HRM) (beamspot size 150$\times$20 $\mu\textrm{m}^2$) and nested HRM (beamspot size 20$\times$20 $\mu\textrm{m}^2$), yielding a combined resolution of 77 meV and 96 meV, respectively.\cite{SuppleMat} Figure \ref{fig:Fig1}(b) shows the experimental geometry. The incoming beam is $\pi$-polarized and the polarization of the outgoing beam collected at a scattering angle of 90$^\circ$ was not analyzed. Thin layers of \ce{RuCl3} were mechanically exfoliated from bulk crystals onto \ce{Si/SiO2} substrates and the selected nanolayers were protected by a thick hBN flake. The substrate was then flushed by oxygen plasma to get rid of unwanted \ce{RuCl3} pieces. Several silver lines pointing at the target flakes were drawn on the substrate surface to facilitate sample alignment in the RIXS chamber. \cite{SuppleMat}

Before presenting the experimental results, we briefly summarize the outcome of previous RIXS experiments on bulk RuCl$_3$ [Fig.1(c)]. The excitation spectrum of interest comprises two segments at low and high energy, respectively: spin-orbit excitations from the $\tilde{S}=1/2$ ground state of the $\textrm{Ru}^{3+}$ ions (electron configuration $d^5$) into the $\tilde{S}=3/2$ excited-state manifold; and {\it d-d} excitations from the $t_{2g}$ crystal-field ground state into the $e_g$ excited states of the Ru ions, which are superposed by a continuum of charge-transfer excitations. Excitations within the $\tilde{S}=1/2$ manifold, which are heavily overdamped in the paramagnetic state, were not studied.

\begin{figure}[!htb]
	\centering
	\includegraphics[width=\columnwidth]{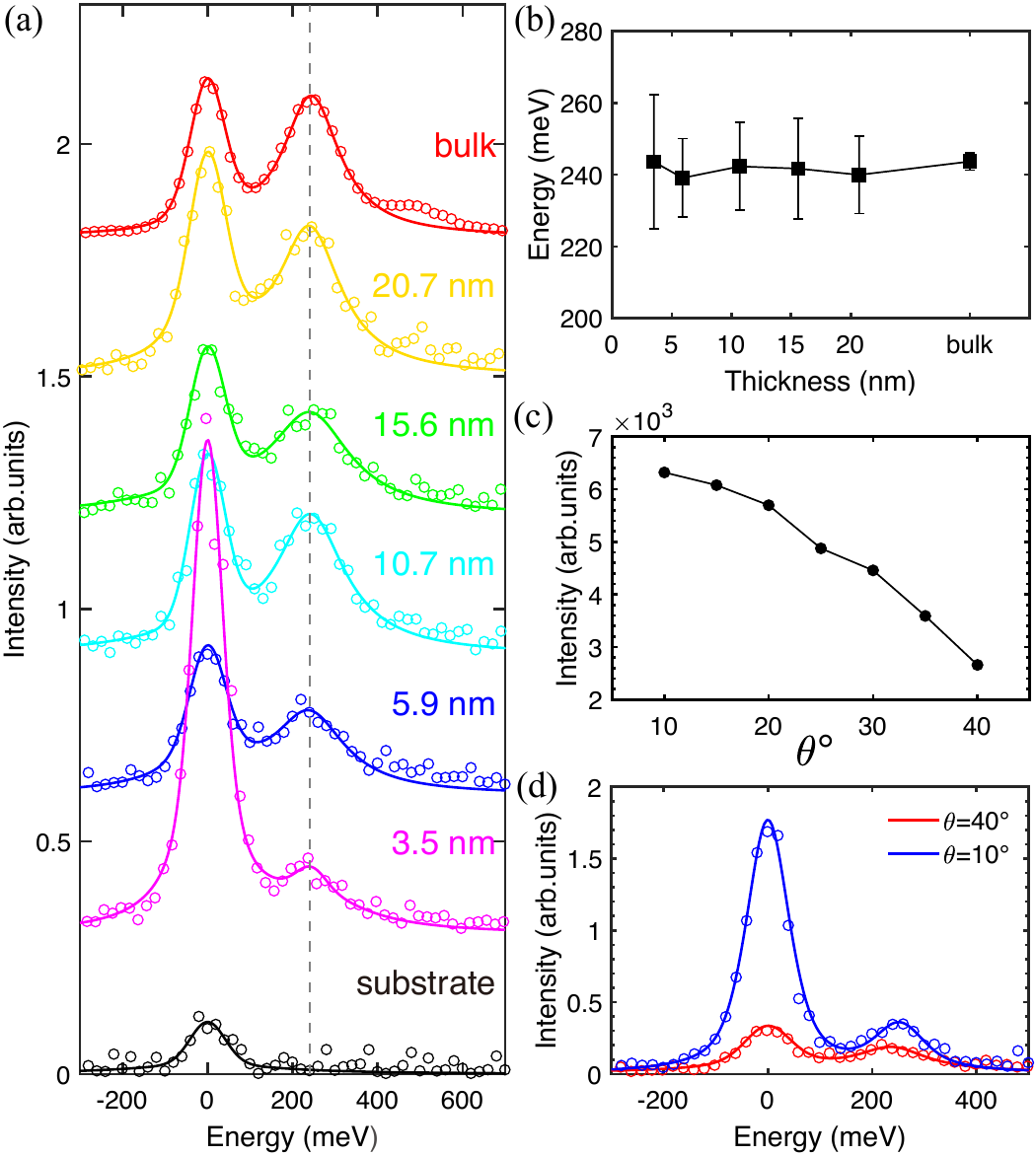}
	\caption{(a) Low-energy RIXS spectra of \ce{RuCl3} nanoflakes, and reference spectrum of a bulk crystal. The incoming X-ray energy was 2837 eV and the sample angle $\theta$=40$^{\circ}$, with in-plane momentum transfer close to the $\Gamma$ point. Details of counting time and HRM configuration are in ref.\cite{SuppleMat}. The spectral intensity of bulk crystal is scaled by a factor of 0.01. Vertical offsets were applied for clarity.
		The grey dashed line is a guide to the eye to indicate the center of the excitation peak. 
		(b) Spin-orbit exciton energies for \ce{RuCl3} bulk crystal and thin flakes. Within the fitting error, the spin-orbit exciton exhibits no thickness-dependent energy shift. (c) Ru-$L_3$ scattering intensity of the 5.9 nm thin flake increases monotonically when approaching grazing-incidence geometry. (d) Low-energy spectra of the 5.9 nm flake at $\theta$=10$^{\circ}$ and 40$^{\circ}$. The spin-orbit exciton peak intensity is enhanced for $\theta$=10$^{\circ}$, despite the large lateral waste of photon flux.}
	\label{fig:Fig2}
\end{figure}

Figure \ref{fig:Fig2}(a) shows the measured low-energy RIXS spectra of nanoflakes with various thicknesses, as well as reference spectra of a \ce{RuCl3} bulk crystal and a \ce{Si/SiO2} substrate. For all measured nanoflakes, we observe an elastic peak due to residual defects in the substrates and samples, and a pronounced inelastic feature around 240 meV. As shown in Fig.\ref{fig:Fig2}(d), the peak energy is almost independent of the incident angle $\theta$ [which modulates the momentum transfer in the honeycomb layers; Fig.\ref{fig:Fig1}(b)]. The lack of a significant momentum-space dispersion implies that this feature arises from a local, intra-atomic excitation. Following prior RIXS studies on bulk RuCl$_3$, \cite{Suzuki2021} we assign it to $\tilde{S}=1/2 \to 3/2$ transitions with energy $\sim 3/2\lambda$, where $\lambda$ is the SOC constant of Ru. Figure \ref{fig:Fig2}(b) shows that the spin-orbit exciton energy is  independent of thickness and identical to the one in bulk crystals. This is expected because the SOC is an intra-atomic interaction that is not significantly influenced by the crystalline environment. Remarkably, despite the increasing lateral photon flux waste at grazing incidence angles due to the small size of the nanoflakes, the RIXS signal increases [Fig.\ref{fig:Fig2}(c,d)], due to the longer travel path within the sample which enhances the scattering probability. \cite{SuppleMat}

\begin{figure}[!htb]
	\centering
	\includegraphics[width=\columnwidth]{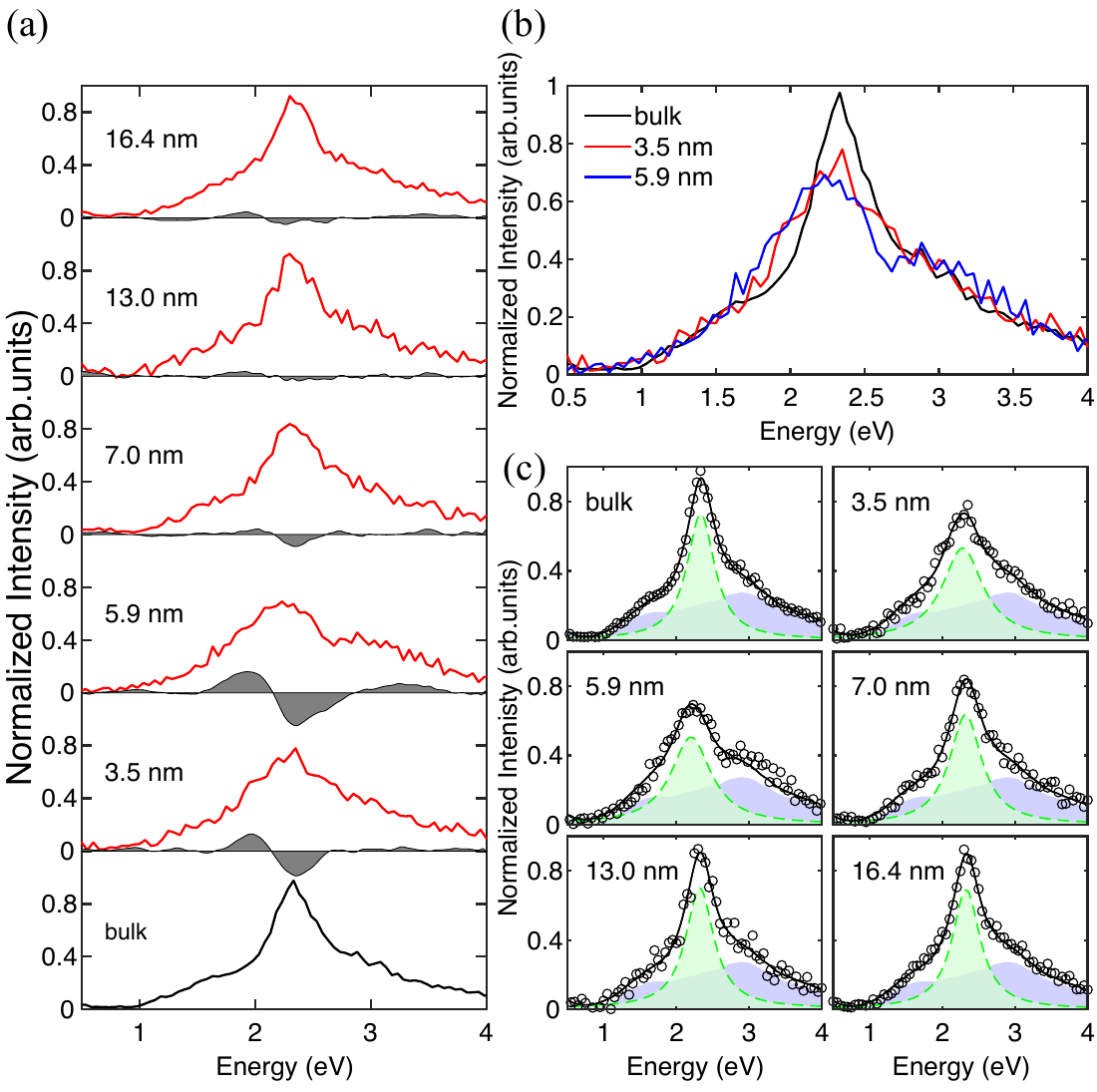}
	\caption{(a) Thickness-dependent multiplet excitation spectra of \ce{RuCl3} for $\theta$=40$^{\circ}$. The incident X-ray energy was 2839 eV.The spectral difference $I_{\textrm{flake}}-I_{\textrm{bulk}}$ (smoothed for clarity) is shown as a grey area. As the flake thickness decreases below 7 nm, a red-shift is observed. (b) Comparison of the spectra of a bulk crystal and two thin flakes. The charge-transfer continuum exhibits no thickness dependent behavior, as seen within the spectral ranges below 1.5 eV and above 3.5 eV. (c) {\it d-d} excitation decomposition for all measured flakes and bulk sample. The blue shaded component represents the charge continuum, independent of flake thickness. The green shaded component is the Lorentzian profile of the main {\it d-d} excitation. Empty circles and black solid lines represent the experimental data and the results of fits to a model function including both components, respectively.}
	\label{fig:Fig3}
\end{figure}

\begin{figure}[!htb]
	\centering
	\includegraphics[width=\columnwidth]{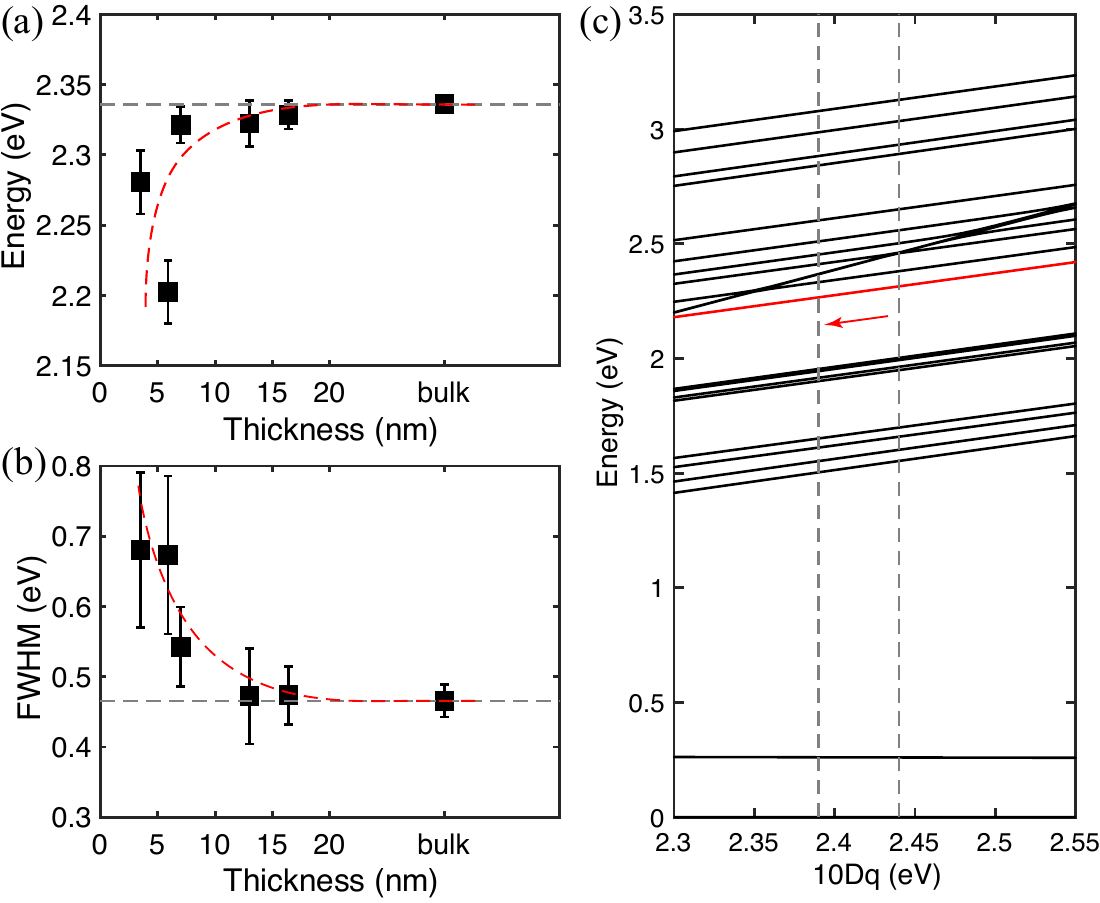}
	\caption{(a) Peak energy and (b) FWHM of the main {\it d-d} excitation resulting from fits. The horizontal dashed lines indicate the values of bulk \ce{RuCl3}. The red curve is a guide to the eye. (c) {\it d-d} excitation energies as a function of the octahedral crystal-field energy $10Dq$ resulting from model calculations. The red line corresponds to the $t_{2g}^4e_g^1$ state that yields the most intense ligand field excitation in the RIXS spectra. The grey vertical lines indicate the bulk value and the average value for the 3.5 nm flake.}
	\label{fig:Fig4}
\end{figure}

The high-energy range of the RIXS spectra comprises a broad inter-site charge-transfer continuum emerging above the charge gap at 1 eV, and sharp {\it d-d} excitation peaks corresponding to intra-ionic crystal-field transitions from the $t^5_{2g}$ ground state to $t^4_{2g}e^1_g$ excited-state multiplets [Fig.\ref{fig:Fig1}(c)]. In agreement with a previous report on bulk RuCl$_3$, \cite{Suzuki2021} we find that a single peak at 2.3 eV dominates the spectrum, whereas other {\it d-d} excitations are much weaker and cannot be clearly separated from the continuum. Figure \ref{fig:Fig3}(a) displays the thickness evolution of the high-energy spectra (normalized to the integrated spectral weight between 1 eV and 4 eV) in comparison to the bulk. The spectral difference $I_{\textrm{flake}}-I_{\textrm{bulk}}$ [grey shaded area in Fig.\ref{fig:Fig3}(a)] calculated from flakes of thickness 7 nm and larger exhibits only minor differences to the bulk. As the thickness decreases further, however, the spectral weight broadens and redistributes towards lower energies. Figure \ref{fig:Fig3}(b) shows a direct comparison between the spectra of bulk \ce{RuCl3} and the two thinnest flakes. The good match in the spectral ranges below 1.5 eV and above 3.5 eV indicates an essentially unchanged charge continuum, and that the observed broadening and red-shift can be mostly ascribed to the main {\it d-d} excitation peak at 2.3 eV. Next the spectra is fitted to a model composed of two components: a Lorentzian profile with variable energy and width describing the main {\it d-d} excitation, and a broad background describing the charge continuum (with submerged minor {\it d-d} excitations) that was kept fixed for all samples.\cite{SuppleMat} The excellent agreement of the resulting profiles with the experimental data [Fig.\ref{fig:Fig3}(c)] indicates that the thickness dependence of the {\it d-d} excitations can be reliably determined by this procedure. Figures \ref{fig:Fig4}(a),(b) show the thickness evolution of the energy and width of the main {\it d-d} excitation profile resulting from these fits. In the two thinnest flakes, the profile is red-shifted by 50-100 meV, and its width increases by about 50\%. 

To clarify the origin of this observation, we implemented a single-ion model calculation based on a Hamiltonian comprising the intra-ionic Hund's coupling and spin-orbit coupling, as well as octahedral and tetragonal crystal fields.\cite{SuppleMat} This method has been widely implemented in RIXS studies to understand and assign the various spectral features and to extract the interaction parameters \cite{Suzuki2021,CaRuO_Hlynur_RIXS_PRB,Hakuto_Takahashi_K2RuCl6_PRL}. We varied each of these parameters while keeping the others fixed at the value of bulk \ce{RuCl3}, and monitored the resulting energy shift of the {\it d-d} feature. The results show that only a shift of the average octahedral crystal field splitting $10Dq$ from 2.44 eV to 2.39 eV can explain the observed red-shift. Varying any of the other parameters within a physically reasonable range does not reproduce the experimental findings.\cite{SuppleMat} 

In a point-charge crystal field model, $10Dq$ is proportional to $1/a^5$, where $a$ is the Ru-Cl bond length, so that the observed red-shift corresponds to an average expansion of the \ce{RuCl6} octahedra by 0.4\%. The concomitant broadening and the systematic thickness evolution of both lineshape parameters [Figs.\ref{fig:Fig4}(a),(b)] imply that any lattice distortion associated with the altered ligand field is inhomogeneously distributed in the out-of-plane direction. We can hence rule out defects or impurities in the RuCl$_3$ crystals from which the flakes were exfoliated (which would give rise to thickness-independent broadening), and bending distortions generated by the exfoliation procedure (which would broaden -- but not shift -- the spectral features from both spin-orbit and crystal-field excitations). Rather, the data point to a mixture of bulk-like inner layers and near-surface layers with different ligand field and, likely, octahedral distortions, which comprise a progressivley larger fraction of the nanoflake volume with decreasing thickness (e.g., 4 inner and 2 surface monolayers in the 3.5 nm sample).
We note that an analogous broadening and red-shift of a peak arising from Cu $d_{x^2-y^2}-d_{3z^2-r^2}$ excitations was observed in a Cu-$L_3$ edge RIXS study of \ce{(CaCuO2)_3}/\ce{(SrTiO3)_2} superlattices, and attributed to the modified crystal structure at the interfaces. \cite{CaCuO2_Minola} We thus conclude that distortions of the RuCl$_6$ octahedra at or near the surface are responsible for the thickness evolution of the crystal-field excitations in our RuCl$_3$ nanoflakes. A survey of the relevant literature has revealed two possible origins of near-surface lattice disorder. First, a theoretical study of RuCl$_3$-based VdW heterostructures  \cite{RuCl3_Gr_strain_PRL} suggests significant strain effects due to lattice mismatch, despite the weak VdW interlayer coupling. By analogy, epitaxial strain at the interface between our RuCl$_3$ flakes and the protective hBN capping layer might increase the Ru-Ru and Ru-Cl bond lengths, and thus weaken the ligand-field interactions. Another possible cause of near-surface lattice distortions are defects such as Cl vacancies, surface adsorbates, or combinations thereof, which are hard to avoid during sample preparation. Evidence of Cl positions different from those in the bulk has indeed been reported in several surface-sensitive experimental studies \cite{RuCl3_surface_distortion_natcomm,RuCl3_surface_distortion_2Dmaterials}, but no agreement has been reached on the nature and strength of these distortions. Our RIXS data can serve as a guide for realistic model calculations of intrinsic and extrinsic lattice distortions and their possible impact on the electronic properties.

In conclusion, we have collected Ru-$L_3$ RIXS spectra of exfoliated \ce{RuCl3} layers with thickness down to 3.5 nm. Although the samples are protected by thick hBN capping layers, and their volumes are orders of magnitude smaller than those of bulk crystals, the signal-to-noise ratio of the RIXS data is sufficient to capture the main spectral features observed in the bulk. We note that all RIXS spectra presented in this work shows no sign of X-ray beam damage [38]. The results reveal a distinct thickness evolution of the low-energy spin-orbit exciton and high-energy crystal-field excitations. Whereas the spin-orbit exciton arises from intra-atomic SOC interactions and is thus independent of thickness, the main crystal-field excitation exhibits a clear broadening and red-shift compared to the bulk, which we are able to attribute to near-surface alternations of the Ru ligand field. Modifications of the Ru-Cl bond lengths and bond angles of the RuCl$_6$ octahedra are important specifically for RuCl$_3$, as they determine the ratio of Kitaev and Heisenberg interactions and hence the propensity for spin-liquid physics. More generally, direct detection of {\it d-d} excitations by RIXS yields insights into the local coordination of transition metal ions and associated ligand fields, which are often hard to access by other spectroscopic methods and can be crucial to the physics of 2D materials and VdW heterostructures, as exemplified by the influence of ligand-field interactions and charge-transfer transitions on the optoelectronic response of atomically thin \ce{CrI3} \cite{CrI3_ligand_PL_naturephy_2018}. Unlike surface-sensitive methods, RIXS is able to detect manifestations of such distortions in samples protected by capping layers, which are routinely used for chemically sensitive 2D materials, and at buried interfaces in VdW heterostructures.

Our results point out various perspectives for further development of the methodology and scope of RIXS experiments on 2D materials. In particular, optimizing the lateral sample dimensions and the experimental geometry (including focusing conditions, incidence and exit angles, background suppression, and acquisition times) should enable measurements on thinner samples, including monolayers and monolayer-based heterostructures. As the energy of the spin-orbit exciton in RuCl$_3$ is comparable to the magnon and paramagnon energies in various tansition metal compounds (including cuprates, iridates, and ruthenates), RIXS experiments on collective spin excitations in 2D materials will also be feasible. Recent advances in high-resolution RIXS instrumentation in the soft, intermediate, and hard x-ray regimes will greatly expand its range of applicability. With these developments, RIXS is poised to realize its potential as a unique source of information on the strength and range of electron-electron and electron-lattice interactions in 2D materials and heterostructures.

\nocite{RuCl3_growth_CVT,PPC_transfer_science,Suzuki2021,Tanabe_Sugano_book,Hund_coupling_AnnRev,Tanabe-Suganoratio,Quanty1,Quanty2,Quanty3}

\begin{acknowledgments}
We thank Giniyat Khaliullin, Huimei Liu, Hakuto Suzuki, and Ziliang Ye for enlightening discussions. We thank Thomas Reindl, Marion Hagel, Ulrike Waizmann and J\"urgen Weis for assistance with sample preparation and Kathrin K\"uster for assistance with AFM measurements. L.W. acknowledges financial support from the Alexander von Humboldt Foundation. We acknowledge DESY (Hamburg, Germany), a member of the Helmholtz Association HGF, for the provision of experimental facilities. The RIXS experiments were carried out at the beamline P01 of PETRA III at DESY. The project was supported by the European Research Council under Advanced Grant No. 669550 (Com4Com).
\end{acknowledgments}

%

\end{document}